\documentclass[aps,prl,showpacs,amssymb, nobibnotes,twocolumn]{revtex4-1}

\usepackage[dvips]{epsfig}
\usepackage{xcolor,soul}
\usepackage{graphics,graphicx}
\usepackage{epstopdf}
\DeclareGraphicsExtensions{.jpg,.pdf,.mps,.png}

\begin{document}

\title{Experimental demonstration of an efficient number diagnostic for long 1D ion chains}

\author{M.R. Kamsap}
\author{C. Champenois}
\email{caroline.champenois@univ-amu.fr}
\author{J. Pedregosa-Gutierrez}
\author{M. Houssin}
\author{M. Knoop}

\affiliation{Aix-Marseille Universit\'e, CNRS, PIIM, UMR 7345, 13397 Marseille, France}

\date{\today}

\begin{abstract}
Very long, one-dimensional (1D) ion chains are the basis for many applications, in particular in quantum information processing and reliable diagnostics are needed to quantify them. To that purpose, we have experimentally validated Dubin's model for very long ion chains [Phys. Rev. Lett. 71, 2753 (1993)]. This diagnostic allows to precisely determine the number of trapped ion with an accuracy of almost 1\% without counting them, by measuring the ion-ion distance of the innermost particles, as well as the trapping potential along the ion chain direction. In our experiment, based on a 155 ion chain, the central 30 ions are measured to be equidistant to better than 2\%,  and we can determine the total number of trapped ions with a 4.5\% uncertainty, completely dominated by a conservative estimation of the experimental characterisation of the trap. 
 \end{abstract}


\maketitle

Trapped one-dimensional ion chains encounter increasing interest in different areas of modern physics, as they constitute  controllable model systems to test fundamental  questions. They are among the most attractive candidates in quantum information \cite{lin09,li13}, and they  serve as a model system for the study of nano-friction \cite{garcia-mata07,benassi11}, or investigations of the Kibble-Zurek mechanism \cite{del_campo10,pyka13,ulm13}. One dimensional strings are also of interest in material science for the study of systems of reduced dimensionality \cite{senga14}, and in colloidal systems where they are used to study non-equilibrium behaviour \cite{straube11}.

Regarding trapped ions as support for quantum computation, the key challenge now is to increase the number of trapped ion qubits to a level where simulations that are intractable otherwise 
might become possible\cite{blatt08}. One  strategy pursued for up-scaling  is to operate with small linear ion chains and multiplex the system by shuttling ions between multiple chains, which requires a dedicated electrode structure and very precise control of ion trajectories \cite{kielpinski02}. Another approach is based on a long linear trap in order to store a single 1D ion-chain containing more than a hundred ions \cite{lin09}. One of the issues raised by this approach is the structural stability of the chain, which requires the aspect ratio of the trapping potential to fulfil a condition which depends on the number of trapped ions \cite{dubin93b, steane97}. It then becomes mandatory to have an efficient diagnostic of the number of trapped ions, in a regime where simply counting them on a recorded image, like routinely done for up to 50 ions, may not be feasible anymore. Furthermore, it would be useful to have a global criterion  linking the minimum number of ions to the local homogeneity at the chain center, opening the way for the implementation of dedicated quantum information protocols \cite{lin09}.

In this letter, we propose an efficient diagnostic for the total number of ions in a chain, even if only the central part of the chain can be recorded. This diagnostic does not rely on the individual counting of each ion but on the measurement of the average inter-ion distance at the center of the chain.   To that purpose, we experimentally validate the model developed by  Dubin in 1993 \cite{dubin93b}, which allows one to deduce the total number of ions in the trap with an ultimate precision of a few percent. This measurement is also a diagnostic for the local inhomogeneity of the chain.

A one-dimensional chain of identical, charged particles  trapped 
along a single  axis can  be experimentally created in a linear radiofrequency (rf) trap. 
  In most linear rf traps, the trapping potential along the chain axis can be very well approximated by a harmonic potential and the equilibrium position of each particle along the trap axis results from the balance between the Coulomb interactions with all the other particles and the restoring trapping force. In many experiments, the trapped ions are laser-cooled to temperatures below 10~mK and they make small oscillations around their equilibrium position. They are typically separated by a distance of a few $\mu$m enabling individual detection of their fluorescence and/or individual addressing by a tightly focussed laser beam. The observation of ion positions for chains longer than a hundred particles allows to experimentally study a correlated system where the correlation energy between particles is as important as the mean-field energy \cite{dubin97}. This situation is rarely met, and for a 3D-system, it takes a $1/r^3$ interaction potential to fulfill this condition. The strong role of particle correlation in the equilibrium energy makes analytical calculations complicated. D.H.E. Dubin takes into account these correlations through the Local Density Approximation and is able to describe the dependency of the local density with the chain length and the total number of particles\cite{dubin93, dubin97}. Once confirmed, these theoretical laws can be used as an efficient diagnostic for the particle number and local homogeneity  for long ion chains, suppressing the need for  individual counting of the ions.

To validate these laws, our experimental results are based on the imaging of the laser induced fluorescence of individual ions on a intensified CCD camera. Long ion chains are routinely created in several groups and parts of ion chains have been observed and reported before (e.g. \cite{enzer00,pyka13}) but they have been limited to ion numbers below 40 mostly because the collecting optics set-up is not suited for translation along the chain. The technical challenge faced by our measurement is the controlled translation of the imaging optics along the trap axis without destruction of the chain \cite{kamsap15t}, allowing us to record photos along the ion string of 4.6 mm length, in order to build a  picture of the full string.

In this letter, we present the existing theoretical laws for ion string structures, then describe the experimental realisation of a very long ion chain, followed by a discussion of the uncertainty budget of the applied method.


A very prolate potential shape is needed to force  a sample of the order of 100  ions  to organise as a chain.  In a  trapping potential which is harmonic in the three directions of space, the structural phase transitions  of linear Coulomb crystals are controlled by the  number $N$ of trapped particles and by the trapping potential aspect ratio $\rho$, which reduces to the ratio of the axial over radial potential steepness $\rho=\omega_z^2/\omega_r^2$ in the case of a linear trap with cylindrical symmetry  \cite{dubin93b, schiffer93, steane97,morigi04}. In the following, we assume that the stability condition for a linear chain is satisfied.  In \cite{dubin93b}, Dubin treats the $N$ ion chain as a charged fluid at zero temperature with total charge $NQ$ and shows that the local density, or number of ions per unit length along the chain, is given by
\begin{equation}
1/a(z)=\left(\frac{3N}{4L}\left(1-\frac{z^2}{L^2}\right)\right)
\label{eq_dubin_a}
\end{equation}
where $a(z)$ is  the distance between closest neighbours, $z$ is the distance from the chain centre and $2L$ is the chain's total length. This organisation is characterised by an increasing inter-ion distance for larger distance to the chain center. Taking into account correlations between ions, Dubin derived the half-length $L$ in terms of the two-body equilibrium length scale $l=(Q^2/4\pi\epsilon_0 m \omega_z^2)^{1/3}$ by 
\begin{equation}
L=l(3N)^{1/3}(\ln N+\ln 6+\gamma_e-13/5)^{1/3}
\label{eq_dubin_L}
\end{equation}
with $\gamma_e\approx 0.577$ the Euler constant ($m$ is the particle mass). Later work \cite{morigi04} confirmed the scaling law $L \simeq l(3N)^{(1/3)}\ln N$ for $N\gg1$. The combination of Eqs.(\ref{eq_dubin_a}) and (\ref{eq_dubin_L}) allows the smallest inter-ion distance $a_0^D$ to be deduced from the number of ions in the chain and the characteristic length $l$ : 
\begin{equation}
a_0^D =4l(3N)^{-2/3}(\ln N+\ln 6+\gamma_e-13/5)^{1/3}
\label{eq_dubin_a0}
\end{equation}

An alternative tool to study self-organization of ions is molecular dynamics simulations. In \cite{james98}, James  proposed an empirical law for the closest distance $a_0^J$ fitting its molecular dynamics results for $N \le 50$ :
\begin{equation}
a_0^J=2.018 l N^{-0.559}
\label{eq_james_a}
\end{equation}
Both predictions scale with $l$ and $N$, and the validation of the described formalisms, allows one to exploit the measurement of $a_0$ as a diagnostic for ion number. The length and number of ions of any long ion chain can then be computed from the measured closest distance $a_0$ and the  length scale $l$, two parameters which are accessible to almost any experimental set-up even with a spatially limited imaging system. 


Our experimental set-up for trapping a long ion chain is based on a linear radio-frequency trap where the distance between the trap center and the rods polarised by rf-voltage is $r_0=3.93$~mm and the distance between the plane electrodes confining the ions along the trap $z$-axis is 96~mm. The $V_{\mathrm{dc}}=2000$~V applied to the $z$-electrodes results in a harmonic potential of $\omega_z/2\pi=2.95\pm 0.13$~kHz. Ions are created  by isotope selective photo-ionisation of a neutral calcium beam  and are then shuttled to the part of the trap where the long chain is formed \cite{kamsap15t} to avoid perturbation of the potential by any calcium deposit on the electrodes.  Ions are laser-cooled by two collimated laser beams close to resonance with the [$4S_{1/2} - 4P_{1/2}$]-transition, of equal power (397~nm, 2~mW  on a  2~mm $1/e^2$  diameter), counter-propagating along the trap axis. Once excited from the ground state, calcium ions can relax to a long-lived metastable  state [$3D_{3/2}$] from which they have to be re-pumped to maintain efficient laser cooling. This re-pumping process is assured by a 866~nm laser beam [resonant with $3D_{3/2} - 4P_{1/2}$] of approximately 2.5~mW and  4~mm $1/e^2$  diameter which co-propagates with one of the cooling lasers. The trapped ions are observed through their laser induced  fluorescence  on the cooling transition which is collected along a direction perpendicular to the trap symmetry axis, and projected onto a photomultipler (PM)  and an intensified  charge coupled device (CCD) camera.  
 \begin{figure*}[htb]
\begin{center}
\includegraphics[width=20.cm]{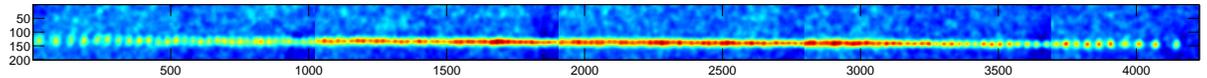}
\caption{ Picture of an ion  chain of 155 ions composed  from 5 pictures taken with a translated objective. Axes are in pixels. }\label{fig_image_chaine}
\end{center}
\end{figure*}
The camera screen is an array of $1024 \times 1024$ pixels with a pixel size of  $13~\mu$m. In order to obtain a sufficient spatial resolution to distinguish ions separated by a few~$\mu$m,  an optical magnification  larger than 10 is required, which limits the number of observable ions to less than 100. These values are representative for many ion-trap experiments. In our set-up, the magnification is 11.58, and the total length of the chain is 4.6~mm, requiring the translation of the  imaging optics  parallel to the trap symmetry axis in order to observe the chain from one end to the other. The objective used for imaging is mounted on a translation stage parallel to this trap axis, and is controlled by a graduated micrometer screw  with graduations of $10~\mu$m. Translations were operated always in the same direction, along the ion chain and result in a $\pm 5~\mu$m uncertainty for every translation  on the position of the chain picture on the camera.

\begin{figure}[htb]
\begin{center}
\includegraphics[width=9.cm]{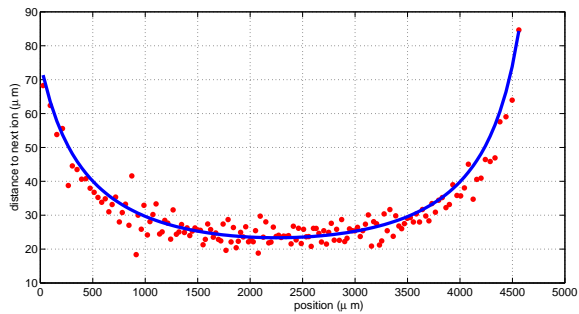}
\caption{Red dots : distance between neighbouring ions $a(z)$ vs their position in the chain, in $\mu$m,  in a potential well  defined by $\omega_z/2\pi=2.95\pm0.13$~kHz. Blue line :  best fit of the inverse of the distance by Eq.~(\ref{eq_dubin_a}) where $z$ is replaced by $z-z_0$ and $N=155$.  }\label{fig_position_chain}
\end{center}
\end{figure}

The translation step of 1~mm is smaller than the camera sensor size in the object space. Some ions are then redundant and are recorded on two consecutive pictures. The average inter-ion distance is 26.7~pxl which is larger than the uncertainty associated to one translation ($\pm 5$~pxl) and which results in 6 redundant ions for each translation (4 for the last one).  These numbers are sufficient to check precisely the consistency of the superposition and deduce an exact number of 155  ions in the chain from  counting the individual ion fluorescence images.  A global picture of the whole chain can be  built, and  is shown on Fig.~(\ref{fig_image_chaine}). The  total length of the chain has been measured to be $2L_m=4619.6~\mu$m. The set of  nearest neighbour distances $a(z)$ is extracted from the data set obtained by vertical binning of the image, and multiple-Gaussian fit to the resulting fluorescence profile. The resulting $a(z)$ are represented on Fig.~(\ref{fig_position_chain}).

We have fitted these data with Eq.~(\ref{eq_dubin_a}) fixing $N$  to 155. The best fit finds a chain center $z_0$ located at the physical center of the chain but a chain length ($2L_{f}=5432.4~\mu$m) which is more than 17\% larger than the measured physical length $2L_m$. If we use  Eq.~(\ref{eq_dubin_L}) for connecting the measured length and number value  $L_m$ and $N$ to the trap characteristic, we obtain a $\omega_z/2\pi=3.48$~kHz. This value  is several $\sigma$ higher than the measured $\omega_z$, whose uncertainty will be discussed later in this article.

We can further check the consistency of the model proposed by Dubin by comparing the inter-ion distance at the center of the chain $a_0$ with the one given by Eq.~(\ref{eq_dubin_a0}) which shows no explicit dependence on $L$. To reduce the experimental uncertainty on $a_0$, we average $a(z)$ for $N_a$ central ions. This number $N_a$ can not be extended too far as there is a quadratic dispersion in inter-ion distance with the distance from the center. If we use  Eq.~(\ref{eq_dubin_a}) around the chain center, one can show that the dispersion in the distance to next neighbour is
\begin{equation}
\frac{\delta a}{a}=\left(\frac{2 N_a}{3 N}\right)^2.
\label{eq_fluctu_a0}
\end{equation}
Considering the dispersion of our experimental data, the best compromise for which the variance reduces to the minimal dispersion one can expect by Eq.~(\ref{eq_fluctu_a0}) is reached for $N_a=30$. The experimental average is then $\overline{a_0}=24.1 \pm 0.2~\mu$m. Assuming for $l$ the value deduced from the direct measurement $\omega_z/2\pi=2.95\pm0.13$~kHz, we can deduce for the number $N_D$ of ions in the chain from Eq.~(\ref{eq_dubin_a0}), $N_D=157 \pm 8$ where the error bar of $\pm$5\% is dominated by the uncertainty on $\omega_z$. The accuracy of the prediction of Eq.~(\ref{eq_dubin_a0}) can be estimated to approximately 1\%. With the same measured parameters, the empirical law proposed by James overestimates the ion number to $N_J=177 \pm 8$.  

To be fully convincing, we have identified the bias that could mislead the experiments to a wrong $N$ estimation from Eq.~(\ref{eq_dubin_a0}). As this estimation is based on a comparison of the two lengths $l$ and $a_0$, the experimental bias could come from the optical magnification, which is needed to extract $a_0$ from the ion fluorescence  image,  and the characterisation of the local potential by $\omega_z$, which defines $l$.

The optical magnification is measured at the chain center, by 8  successive translations of the objective, perpendicular to the optical axis and to the chain axis, over a total range of 1~mm. These translations are driven by a $10~\mu$m resolution screw and their linear correlation to the chain picture position on the camera allows us to estimate the 11.58 magnification mentioned earlier in the text. The uncertainty induced by the mechanical precision is an order of magnitude lower than the magnification shift induced by the translation of the objective along the trap axis to capture the full chain, calculated to be 0.028. The uncertainty induced on $\overline{a_0}$ is one order of magnitude smaller than the one due to the data dispersion, and thus does not play a major role here.

The measurement of the axial frequency of motion of the ions in the trap, $\omega_z$,  was done by observing the fluorescence drop which is induced when a small additional oscillating potential applied on one of the DC electrodes is in resonance with  a frequency of motion  \cite{champenois01}.  To distinguish the axial motional frequency from the radial ones and reach an increased precision in our measurement, we follow the resonance lines which shift in frequency with $V_{\mathrm{dc}}$ (scaling in $\sqrt{V_{\mathrm{dc}}}$). Due to screening and the extreme length of our trap, $\omega_z$ is  equal to 2.95~kHz when we apply the maximum $V_{\mathrm{dc}}$  of 2000~V. The precision deduced from a fit to the $V_{\mathrm{dc}}$ dependence is $\pm 0.13$~kHz and is due to the spectral width of the fluorescence dip.  The relative uncertainty is large because the absolute value of $\omega_z$ is small compared to usual oscillation frequency in rf trap. The same technique can be used to identify the radial frequencies of motion, by looking at resonances whose frequency depends on $V_{\mathrm{rf}}$, the time oscillating voltage amplitude, applied between adjacent  rods. We find in this case $\omega_x/2\pi=(157\pm 1) \times V_{\mathrm{rf}}/2000$~kHz.

\begin{figure}[htb]
\begin{center}
\includegraphics[width=8.cm]{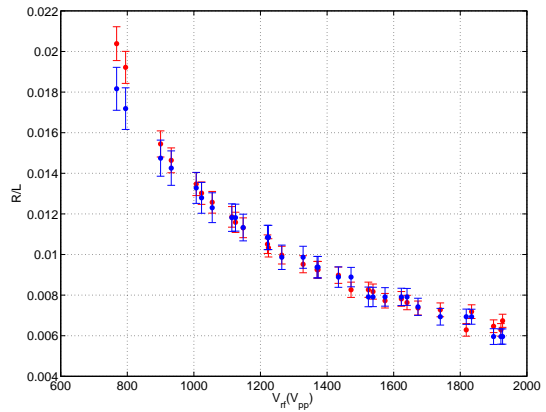}
\caption{Aspect ratios of a cloud versus the rf amplitude $V_{\mathrm{rf}}$ for $V_{\mathrm{dc}}=2000$~V, either directly measured on its picture (red dots) or calculated from $\omega_x/2\pi=(157\pm 1) \times V_{\mathrm{rf}}/2000$~kHz and $\omega_z=2.95 \pm 0.13$~kHz (blue dots).}\label{fig_aspect_ratio}
\end{center}
\end{figure}

 The accuracy and consistency of the measured frequency values can be tested by an independent  method based on  the observation, in the same trap,  of large clouds of typically few thousand particles cooled down to the liquid phase. The sample is then characterised by a uniform density, depending only on the values of the rf-induced potential like $\omega_x^2 \propto V_{\mathrm{rf}}^2$ \cite{dubin99}.  The trapped and laser-cooled ions form an ellipse whose aspect ratio $\alpha=R/L$ depends solely on the aspect ratio of the potential $\omega_z^2/\omega_r^2$ \cite{turner87} where $\omega_r^2=\omega_x^2-\omega_z^2/2$. For a given cloud of 12.5~mm length, we have measured the aspect ratio for different rf-amplitudes $V_{\mathrm{rf}}$. Due to the very small aspect ratio of the trapping potential, the cloud is very elongated and its aspect ratio is smaller than 0.01. Its measurement for different rf-amplitudes $V_{\mathrm{rf}}$ can be done without moving the objective along the cloud because its length is nearly constant. In the limit of an aspect ratio far smaller than 1, one can show that the dependence of $\alpha$ with  $\omega_z\omega_r$ is nearly linear and,  for a given number of ions, $\delta L/\delta V_{\mathrm{rf}} \simeq 0$. The volume modification takes place on the radial dimension of the cloud and $\delta R/R\simeq -\delta\omega_x/\omega_x$. As a consequence, the measurement of the cloud aspect ratio for different $\omega_x$  requires to measure the length of the cloud only once, and to measure the radial size for each $V_{\mathrm{rf}}$. The results are compared to the values calculated from the values of $\omega_z$ and $\omega_r$ measured by fluorescence dip technique, they are both plotted on Fig.~\ref{fig_aspect_ratio}. The measured and calculated aspect ratios match very well within the error bar, confirming the appropriate description of the trapping potential. 

In conclusion,  even if the total length involved in the equations is not in agreement with the experiment, we have experimentally validated the local description derived for long 1-D Coulomb system in a harmonic potential, an example of a strongly correlated system. This law is very useful as a diagnostic to prepare a chain with a sufficiently large number of ions to have a minimum number of equidistant ions in its central part. The total number of ions can be derived  without individually counting them, if the potential well along the chain direction is well known. With the described  155-ion chain, we were able to retrieve the ion number from a single composed picture with an accuracy of approximately 1\% and an uncertainty of $\pm 4.5\%$ completely determined by the measurement uncertainty on the axial potential. The 30 central  ions  are equidistant to better than 2\%. Such a system could be the starting point for explorations of quantum information protocols  as proposed in \cite{lin09}. The validation of this straightforward  number diagnostic which does not require fastidious   counting of individual ions, makes these systems simpler to control, and therefore more attractive.

Fruitful discussions with J\'er\^ome Daligault and Dominique Escande are gratefully acknowledged. This experiment has been financially supported by ANR (ANR-08-JCJC-0053-01), CNES (contract 116279) and R\'egion PACA. MRK acknowledges financial support from CNES and R\'egion Provence-Alpes-Cote d'Azur.

%

\end{document}